\documentclass[procedia]{easychair}

\usepackage{doc}
\usepackage{makeidx}

\usepackage{graphicx}

\def\procediaConference{99th Conference on Topics of
  Superb Significance (COOL 2014)}

\title{Exact Solutions on the Ground States of Ising Models in  Magnetic Fields with Frustration on a Diamond Hierarchical Lattice}

\titlerunning{Exact Solutions on the Ground States ...}

\author{
    \underline{Yuhei Hirose}
\and
    Akihide Oguchi
\and
   Yoshiyuki Fukumoto\\
}

\institute{
  Tokyo University of Science, Noda, Chiba 278-8510, Japan\\
  \email{yfuku@rs.tus.ac.jp}
 }

\authorrunning{Yuhei Hirose, Akihide Oguchi, and Yoshiyuki Fukumoto}

\begin{document}

\maketitle


\begin{abstract}
Magnetization processes of Ising models with frustration on diamond hierarchical lattices, which contain vertices with high coordination numbers, are exactly obtained at zero temperature. 
In antiferromagnetic systems, the magnetization cannot saturate under finite magnetic fields owing to the competition between the antiferromagnetic and  Zeeman interactions 
and the intrinsic long-range nature of hierarchical lattices. 
For the zero-field classical spin-liquid phase found in [Kobayashi et al., J. Phys. Soc. Jpn. 78, 074004 (2009)], an infinitely small applied magnetic field can induce an infinitely small magnetization, 
despite Ising models that have discrete energy levels.  
By examining the structure of the partition function, we obtain the ground state spin-configurations and clarify the mechanism of the ``gapless like behavior''.
\end{abstract}


%
%


\section{Introduction}
\label{sect:introduction}

Studying the structure of the multistep magnetization is one of the central challenge in frustrated spin systems. The most extreme example, which was shown by Bak and Bruinsma in 1982, is the complete devil's staircase in the magnetization curve of the antiferromagnetic Ising chain with long-range interaction\cite{Bak1982}.

Recently, we have derived exact solution of the system with multistep magnetization\cite{Hirose2014}. 
We consider a frustrated Ising model on diamond hierarchical lattices\cite{Kobayashi2009}. 
The diamond hierarchical lattices are constructed via infinite iteration procedures and have vertices with coordination numbers that increase with increased stage as shown in Fig.1. 
Thus the diamond hierarchical lattices have an inherent long-range nature. 

In our study,  the magnetization processes of Ising models with frustration on diamond hierarchical lattices are exactly obtained at zero temperature. To define the Ising model on each diamond-hierarchical lattice, we place a Ising spins, $\sigma_i=\pm 1$, on each vertex. 
The spins on both ends of the solid line and of the dotted line couple are represented by $\pm J$ ($J>0$) and $\alpha J$ ($\alpha>0$), respectively. 
The Hamiltonian can be written as
\begin{equation} 
   {\cal H}=\pm J\sum_{\langle i,j\rangle} \sigma_i \sigma_j +\alpha J \sum_{\langle\!\langle i,j\rangle\!\rangle}\sigma_i \sigma_j -H\sum_{i}\sigma_{i}
   \label{eq:1},
\end{equation}
where $H$ is the magnetic field strength, the first sum runs over all pairs of nearest neighbors on solid bonds, and the second sum runs over all pairs on dotted bonds. 
The diamond unit, which is the same as the second-stage lattice, has the most frustrated ground state at $\alpha=2$, where the ferro- and antiferromagnetic configurations  have the same energy value.

\begin{figure}[h]
\begin{center}
\includegraphics[width=.95\linewidth]{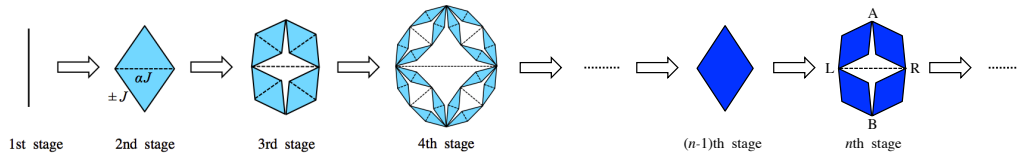}
\end{center}
\caption{
Construction of diamond hierarchical lattice. $+J$ and $-J$ represent the antiferromagnetic and ferromagnetic exchange interaction respectively.
Spins on the sites indicated by A and B are called ``top spins'' and those on the sites indicated by L and R are called ``diagonal spins''.}
\label{Fig1}
\end{figure}

We find that the behavior changes substantially as the value of $\alpha$ is varied and show that there are several types of infinitely multiple-step (IMS) structure in a certain magnetic region. 
In the case of $\alpha>2$, we have the paramagnetic phase and have independent $(\uparrow, \downarrow)$ pairs formed by antiferromagnetic diagonal bonds. 
In the case of $\alpha<2$, for the zero-field classical spin-liquid phase with highly developed short-range order, 
an infinitely small applied magnetic field can induce an infinitely small magnetization despite Ising models that have discrete energy levels. 
By examining the structure of the partition function, we obtain the ground state spin-configurations and clarify the mechanism of the ``gapless like behavior''. 
We also find that, in the antiferromagnetic case, the magnetization cannot be saturated under finite magnetic fields even if the antiferromagnetic diagonal coupling vanishes. 
On the other hand, in the ferromagnetic case, we obtain the saturated magnetization at a finite saturation field.

This paper is organized as follows. 
In sect. 2, our procedure to calculate the ground-state partition function are described. The magnetization and entropy curves are studied in Sect.3. In sect.4, we summarize the results obtained in this study.

\section{Partition Function and Recursion Relations in the\\
 $n$-Stage Lattice}
\label{sect:Hamiltonian and Partition Function}

We can write the partition function at the $n$-stage lattice
\begin{equation} 
Z_n=e^{2L}a_{n}+2b_{n}+e^{-2L}c_{n},
\label{eq:2}
\end{equation}
where $a_n$, $b_n$ and $c_n$ are partition functions under the condition that the top spin state are, respectively, 
fixed to $(\uparrow, \uparrow)$, $(\uparrow, \downarrow)$ or $(\downarrow, \uparrow)$, and $(\downarrow, \downarrow)$. 
We can write the recurrence equations
\begin{align} 
   a_{n}=&e^{2L-B}a_{n-1}^{4}+2e^{B}a_{n-1}^{2}b_{n-1}^{2}+e^{-2L-B}b_{n-1}^{4}
   \label{eq:3},\\
   b_{n}=&e^{2L-B}a_{n-1}^{2}b_{n-1}^{2}+2e^{B}a_{n-1}c_{n-1}b_{n-1}^{2}+e^{-2L-B}b_{n-1}^{2}c_{n-1}^{2}
   \label{eq:4}, \\
   c_{n}=&e^{2L-B}b_{n-1}^{4}+2e^{B}b_{n-1}^{2}c_{n-1}^{2}+e^{-2L-B}c_{n-1}^{4}
   \label{eq:5},
\end{align}
where $a_1=c_1=e^{\mp K}$ and $b_1=e^{\pm K}$ with $K=J/T$, $B=\alpha J/T$, and $L=H/T$. (We use units corresponding to $k_{\rm{B}}=1$ in this paper.) 
The recurrence equations in Eqs.(\ref{eq:3})-(\ref{eq:5}) are shown schematically in Fig.2. 

\begin{figure}[b]
\begin{center}
\includegraphics[width=.70\linewidth]{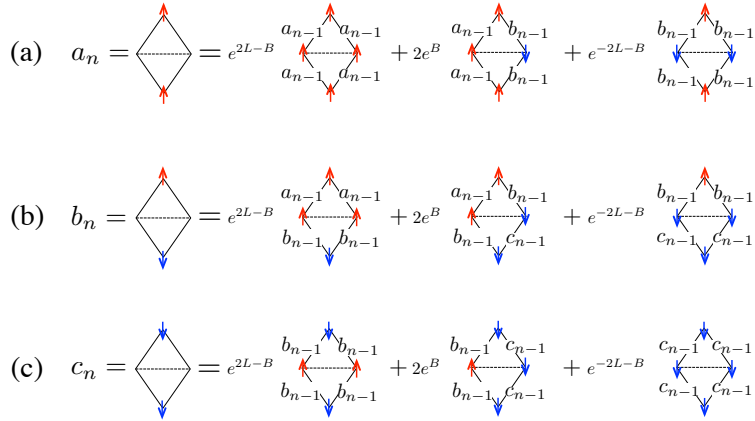}
\end{center}
\caption{
Schematic representations of the recurrence formulas for the conditional partition functions. The $n$-stage lattice is composed of four $(n-1)$-stage lattice, and solid lines represent the $(n-1)$-stage lattice. }
\label{fig:2}
\end{figure}

Obtaining the $n$-stage partition function is equivalent to obtaining $a_{n}$, $b_{n}$, and $c_{n}$ 
by using repeatedly the recursion equations with initial values of $a_{2}$, $b_{2}$ and $c_{2}$, 
but this is very difficult because $n$ is increased, the number of terms appearing in the partition function increases like $3^{4^{n-2}}$. 
Therefore, we restrict ourselves to the absolute zero temperature and only consider the largest term in the right hand sides of the relations given in Eqs.(\ref{eq:3})-(\ref{eq:5}).
By repeating  the recursion formula several times, and choosing the largest term, 
we could find simple functional relations : $b_i=e^{-\lambda}c_i, a_i=e^{-2\lambda}c_i$ where the value of $i$  depends on the strength of the magnetic field and $\lambda$ is function of $K, B$, and $L$ \cite{Hirose2014}.

By using the functional relations, we obtain the $n$-stage partition function. 
We concentrate on the antiferromagnetic case below. (Result for the ferromagnetic case is given by \cite{Hirose2015}.) The partition function is written by
\begin{align}
\log Z_n =& \log e^{-2L}+ 4^{n-i} \log c_i + \sum_{r=1}^{n-i} 4^{n-(i+r)} \log F(2^r \lambda)
\label{eq:6},
\end{align}
where
\begin{align}
F(2^{r}\lambda)=& e^{2L-B-2^{r+1}\lambda}+2e^{B-2^{r}\lambda}+e^{-2L-B},
\label{eq:7}
\end{align}
for $T\to 0$.  
Furthermore, we can obtain the magnetization per spin from
\begin{align}
m=&\lim_{n \to \infty} \frac{1}{N_n}\frac{\partial}{\partial L} \log Z_n,
 \label{eq:8} 
\end{align}
for $m \in [0, 1]$ and the entropy per spin from
\begin{align}
s=&\lim_{n \to \infty} \frac{1}{N_n}\left.\frac{\partial}{\partial T}(T\log Z_n)\right|_{T=0},
 \label{eq:9}
\end{align}
for $s \in [0, \log2]$, where $N_n=\frac{2}{3}(4^{n-1}+2)$ is the site number. The partition function, which is included in Eqs. (\ref{eq:8})-(\ref{eq:9}), is written by Eq. (\ref{eq:6}), and $F(2^{r}\lambda)$ is considered to be the largest term in the right hand side of Eq. (\ref{eq:7}) under the absolute zero temperature  \cite{Hirose2015}.

\section{Magnetization and Entropy}
\label{sect:Magnetization and Entropy}

In this section, we show the dependence of the magnetization and the entropy on magnetic field for the case of $2<\alpha \le 3$ and $\frac{3}{2} \le \alpha<2$ in the antiferromagnet.

\subsection{Case of $2<\alpha \le3$}

\begin{figure}[b]
\begin{center}
\includegraphics[width=.95\linewidth]{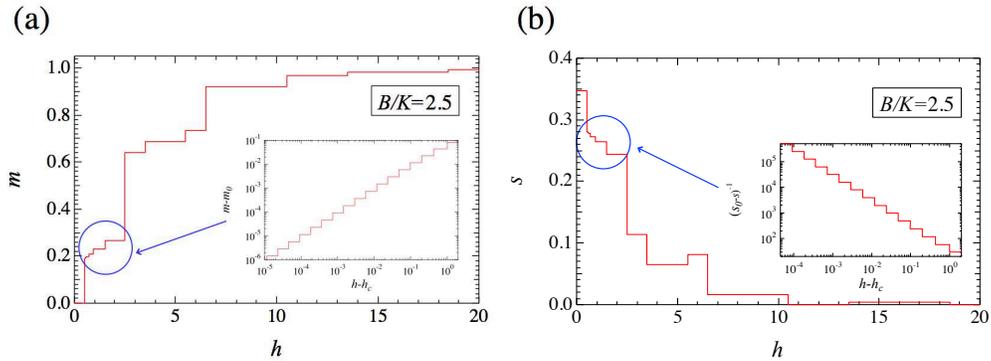}
\caption{Dependence of (a) magnetization per site $m$ and (b) entropy per site $s$ on magnetic field 
$h=H/J=L/K$ for $\alpha=B/K=2.5$.
The inset shows an enlarged plot of region that are circled, where 
$h_{\rm c}=B/K-2=0.5$ is the extreme left of region that are circled. The value of $m_0=\frac{3}{16}$ and $s_0=\frac{13}{32}\log2$ are the magnetization and the entropy at $h=h_{\rm c}+0$ respectively. 
 }
\label{fig:3}
\end{center}
\end{figure}

We discuss the typical features of the resultant magnetization and entropy curve for the case of a strong AF diagonal coupling, $2<\alpha \le 3$. 
We show a calculated result in Fig.~\ref{fig:3}, where $\alpha=B/K=2.5$ was chosen. We have an infinitely large saturation field, i.e., we have the IMS structure around $(h, m)=(\infty, 1)$ and $(h, s)=(\infty, 0)$, where $h \in [0, \infty)$ represents the magnetic field. 
This high-field IMS structure in the antiferromagnet originates from the competition between the  nonfrustrated antiferromagnetic interaction $J$ 
and the magnetic field $H$ and an intrinsic long-range nature  of hierarchical lattices. 
On the other hand, in the ferromagnet, we have the saturated magnetization at a finite saturation field as shown in Figs.~8 and 9 in \cite{Hirose2014}.
In regions that are circled in the Fig.~\ref{fig:3} (a) and (b), other IMS structures appear around $(h, m)=(h_c, \frac{3}{16})$ and $(h, s)=(h_c, \frac{13}{32}\log2)$, where $h_c=\alpha-2$ is a critical field.
In region of $h<h_c$, which is the paramagnetic phase, the magnetization vanishes and the entropy is at a maximum value, $m(h<h_c)=0$ and $s(h<h_c)=\frac{1}{2}\log2$,
because strong antiferromagnetic diagonal bonds form $(\uparrow, \downarrow)$ or $(\downarrow, \uparrow)$ pairs.

\subsection{Case of $\frac{3}{2}\le \alpha<2$}

\begin{figure}[b]
\begin{center}
\includegraphics[width=.95\linewidth]{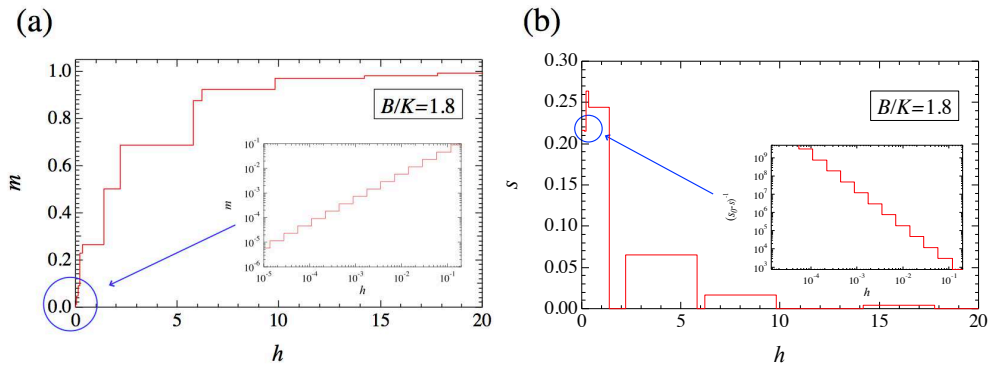}
\caption{Dependence of (a) magnetization per site $m$ and (b) entropy per site $s$ on magnetic field 
$h=H/J=L/K$ for $\alpha=B/K=1.8$.
The inset shows an enlarged plot of region that are circled, where $s_0=\frac{5}{16}\log2$ is the entropy at $h=0$.
 }
\label{fig:4}
\end{center}
\end{figure}

Next, we discuss the typical features of the resultant magnetization and entropy curve for the case of a weak AF diagonal coupling, $\frac{3}{2}\le \alpha<2$. 
The zero-field ground state for $\frac{3}{2}\le \alpha<2$ is the spin liquid state \cite{Kobayashi2009}. 
We show a calculated result in Fig.~\ref{fig:4}, where $\alpha=B/K=1.8$ was chosen.  
We have an infinitely large saturation field or the IMS structure around $(h, m)=(\infty, 1)$ and $(h, s)=(\infty, 0)$, which is the same as in the previous case of $2<\alpha \le3$. 
In regions that are circled in the Fig.~\ref{fig:4} (a) and (b), the IMS structures appear around $(h, m)=(0, 0)$ and $(h, s)=(0, \frac{5}{16}\log2)$ respectively. 
It is interesting to note that an infinitely small magnetic field can induce an infinitely small magnetization, despite Ising models that have discrete energy levels.

\begin{figure}[b]
\begin{center}
\includegraphics[scale=0.7,angle=0]{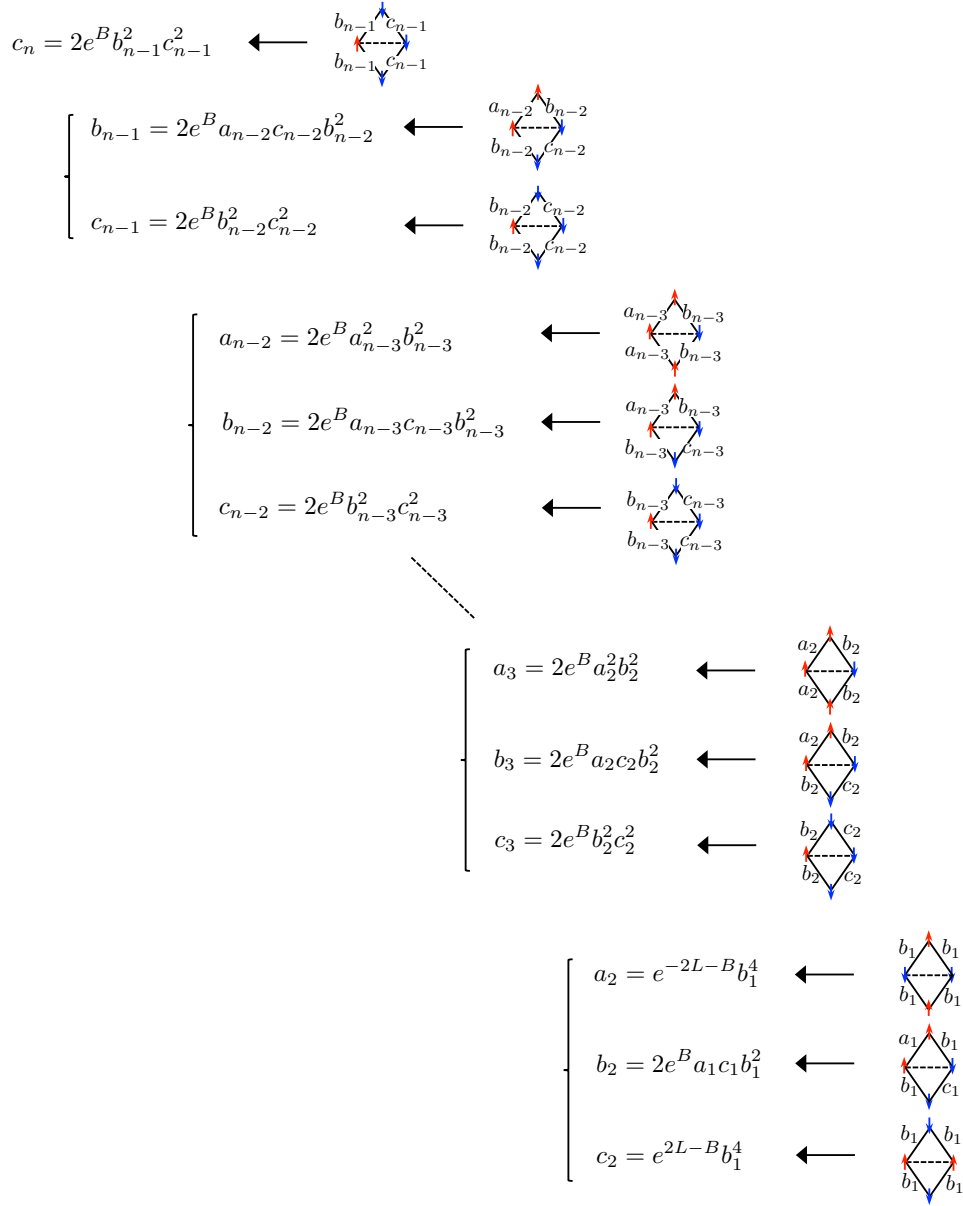}
\end{center}
\caption{Spin configurations for the $n$-stage lattice in the spin-liquid state. The diagonal spins of $a_2$ and $c_2$ are, respectively, $(\downarrow, \downarrow)$ and $(\uparrow, \uparrow)$, 
and those of the other stages are $(\uparrow, \downarrow)$.
}
\label{fig:5}
\end{figure}

\begin{figure}[b]
\begin{center}
\includegraphics[scale=0.5 ,angle=0]{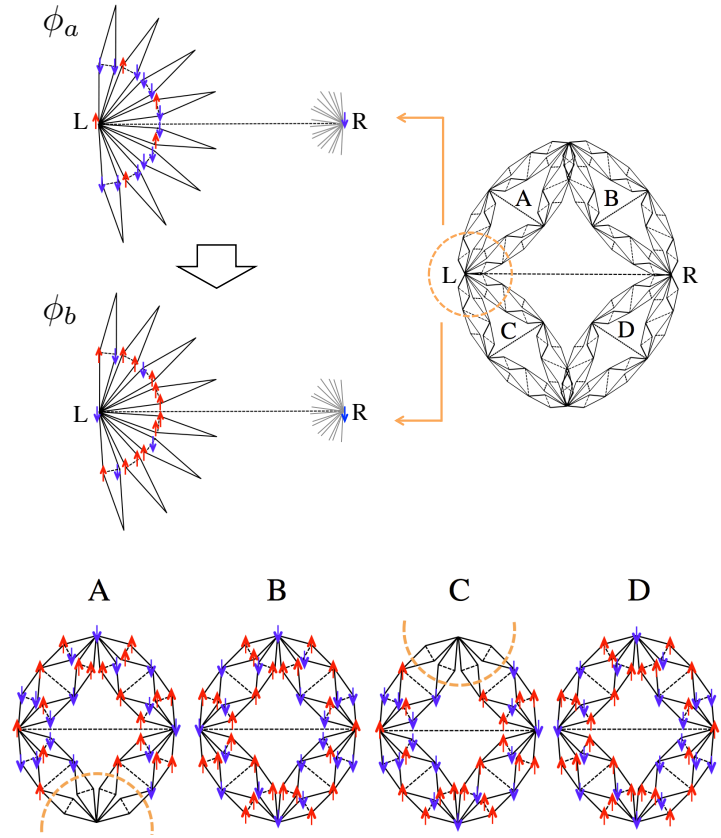}
\end{center}
\caption{Schematic representations of lowest energy spin configurations for $n=5$ : the zero-field state $\phi_a$ and the state $\phi_b$ which destabilizes $\phi_a$ at $L=L_c$. The spins denoted by L and R are diagonal spins of the 5th stage lattice. Spin directions in $\phi_a$ are different from those in $\phi_b$ only for the spins surrounded by the dashed orange circle. For the other spins, their directions are shown in the lower panel, where A, B, C, and D are 4th stage lattices constituting the 5th stage lattice in the upper panel. 
}
\label{fig:6}
\end{figure}

To clarify the mechanism of the above gapless like behabior, which is one of the most interesting phenomena, 
we return to the recursion equations (\ref{eq:3})-(\ref{eq:5}) and the partition function (\ref{eq:6}). 
The coefficient of the partition function (\ref{eq:6}) represents the number of lattices in the $n$-stage lattice.
Thus, by obtaining the spin configurations of conditional partition functions $a_j$, $b_j$, and $c_j$ $(j=1,2,\cdots,n)$ using the recurrence equations, 
and by identifying what kind of spin configurations correspond to each term of the partition function, we can elucidate the lowest-energy spin configuration. 
In Fig.~\ref{fig:5}, we show the spin configurations in the spin liquid state. 
We can see in Fig.~\ref{fig:5} that the diagonal spins of $a_2$ and $c_2$ are, respectively, $(\downarrow, \downarrow)$ and $(\uparrow, \uparrow)$, 
and those of the other stages are $(\uparrow, \downarrow)$.
In other words, the diagonal spins in the second stage contribute to magnetization, but those in the other stages do not.
On the basis of Fig.~\ref{fig:5}, it is possible to enumerate total numbers of $a_2$ and $c_2$ in the $n$-stage lattice:
the total number of $a_2$ is $-2^{n-3}+4^{n-3}$ and that of $c_2$ is $2^{n-3}+4^{n-3}$.
Thus we find the magnetization is given by $M_n=-2(-2^{n-3}+4^{n-3})+2(2^{n-3}+4^{n-3})-2=2^{n-1}-2$, because $a_2$ and $c_2$ has, respectively, magnetizations of $-2$ and $+2$ and the top spins in the $n$-stage lattice has that of $-2$. 
We here consider the case of $n=5$ to show a lowest-energy spin configuration on the real lattice and then describe how a transition of spin configurations is induced by a small magnetic field.
In Fig.~\ref{fig:6}, we show the explicit spin configuration corresponding to that in Fig.~\ref{fig:5}, which is denoted by $\phi_a$.
When the magnetic field is increased, the spin configuration $\phi_a$ is destabilized by the spin configuration $\phi_b$ defined in Fig.~\ref{fig:6}. Denoting the critical field as $L_{\rm c}$, spin flips occur for the diagonal spin L (see the upper panel of Fig.~\ref{fig:6}.) and some of its nearest neighbors connected by the solid lines at $L=L_{\rm c}$.
At this transition, exchange energy loss in diamond units attached to the diagonal spin L cancel each other out
and the net energy loss, $\Delta E_5=2\alpha$, comes from the exchange coupling between the diagonal spins L and R.
(For a general value of $n$, it can be shown that $\Delta E_n=2\alpha$.)
As for change in magnetization, counting the number of diagonal spin pairs with $(\uparrow,\uparrow)$ and $(\downarrow,\downarrow)$ in Fig.~\ref{fig:6}, we have $\Delta M_5=14$. (For a general value of $n$, it can be shown that $\Delta M_n=2^{n-1}-2$.)
Therefore, we obtain a expression for $L_{\rm c}$ as follows
\begin{align}
   L_{\rm c}=\frac{\Delta E_n}{\Delta M_n}=\frac{2\alpha}{2^{n-1}-2}\rightarrow 0\hspace{5mm}(\mbox{for $n\rightarrow \infty$})
   \label{eq:10}
\end{align}
Thus, it is revealed that gapless like behavior originates from the fact that a large number of spins flip upwards following application of an infinitely small magnetic field, keeping the loss of exchange energy finite. 

We discuss why the Ising model on standard lattices does not exhibit such an exotic gapless-like behavior. By examining a transition of spin configurations in the case of $n=5$, we find that spin flips occur for only the diagonal spin L and some of its nearest neighbors connected by the solid lines. This means that the spin with the largest coordination number, and some of its nearest neighbors flip upwards. This holds for the general value $n\ge5$, and but $n\le4$, it does not because of the small coordination number, which indicates that a large coordination number, or a long-range interaction, is an essential role of the gapless-like behavior. Therefore, we can conclude that the appearance of the IMS structure around $(h, m)=(0, 0)$ and that of the gapless-like behavior come from the combination of  an inherent long-range nature of hierarchical lattices and the frustration effect of diamond structures.   

\section{Conclusions}
\label{sect:Conclusions}
Magnetization processes of Ising models with frustration on diamond hierarchical lattices are exactly obtained at absolute zero temperature.  In this paper, we discuss only the antiferromagnetic case.
Our hierarchical lattice has vertices whose coordination numbers increase whenever the stage goes up, so we can regard this behavior as an intrinsic long-range nature, to which the appearance of the IMS structure can be ascribed.

One of the most interesting phenomena is that applying a infinitely small magnetic field on the spin liquid phase gives an infinitely small magnetization; the IMS structure around $(m,h)=(0,0)$ appears. By analyzing the structure of the recursion equations and the partition function, we can elucidate the lowest-energy spin configuration and clarify the origin of the spin-liquid ground state.




\begin{thebibliography}{99}
\bibitem{Bak1982} P. Bak and R. Bruinsma, Phys. Rev. Lett. \textbf{49}, 249 (1982).; R.Bruinsma and P. Bak, Phys. Rev. B \textbf{27}, 5824 (1983).
\bibitem{Hirose2014} Y. Hirose, A. Oguchi, and Y. Fukumoto, J. Phys. Soc. Jpn. \textbf{83}, 074716 (2014).
\bibitem{Kobayashi2009} H. Kobayashi, Y. Fukumoto, and A. Oguchi, J. Phys. Soc. Jpn. \textbf{78}, 074004 (2009).
\bibitem{Hirose2015} Y. Hirose, A. Oguchi, and Y. Fukumoto, J. Phys. Soc. Jpn. \textbf{84}, 104705 (2015).
\end{thebibliography}
\end{document}